\documentclass[sigconf, nonacm]{acmart}

\newcommand\vldbyear{2026}
\newcommand\vldbworkshop{VLDB Ph.D. Workshop}
\newcommand\vldbauthors{\authors}
\newcommand\vldbtitle{Indexing Long Documents for LLM-Based Analysis}
\newcommand\vldbavailabilityurl{}
\newcommand\vldbpagestyle{plain}

\usepackage{tikz}
\usetikzlibrary{calc}

\usepackage{caption}
\definecolor{lin}{HTML}{e31a1c}

\definecolor{donna}{HTML}{1f78b4}

\begin{document}
\title[Indexing Long Documents for LLMs]{Indexing Long Documents for LLM-Based Analysis}

\author{Donna Pham}
\affiliation{%
  \institution{University of Michigan}
  \city{Ann Arbor}
  \state{Michigan}
  \country{USA}
}

\email{phdonn@umich.edu}
\affiliation{%
  \institution{Supervised by Lin Ma, University of Michigan}
}

\begin{abstract}
Long documents such as clinical records, legal contracts, and scientific papers are increasingly analyzed with large language models (LLMs). Naturally, feeding the full document to the model for every question can eventually become slow, expensive, prone to hallucination, and it reuses no work across questions. We explore an indexing-based solution for document analysis and propose a hierarchical plain-text index that is built once per document and consulted by subsequent queries. Inspired by the classic B+ tree, the index organizes a document into pages arranged from general summaries at the root to specific ones at the leaves, but it departs from the B+ tree in three ways suited to LLM access: content lives at every level, the index is plain text rather than attribute values, and navigation follows relevance between page summaries rather than comparing a search key. Its structure is discovered per document by the LLM rather than being hand-designed. In a preliminary evaluation on the NarrativeQA dataset, the index reaches accuracy comparable to DocETL, the strongest baseline, while being 40\% cheaper to answer questions.
\end{abstract}

\maketitle

\pagestyle{\vldbpagestyle}
\begingroup\small\noindent\raggedright\textbf{VLDB Workshop Reference Format:}\\
\vldbauthors. \vldbtitle. VLDB \vldbyear\ Workshop: \vldbworkshop.\\ 
\endgroup
\begingroup
\renewcommand\thefootnote{}\footnote{\noindent
This work is licensed under the Creative Commons BY-NC-ND 4.0 International License. Visit \url{https://creativecommons.org/licenses/by-nc-nd/4.0/} to view a copy of this license. For any use beyond those covered by this license, obtain permission by emailing \href{mailto:info@vldb.org}{info@vldb.org}. Copyright is held by the owner/author(s). Publication rights licensed to the VLDB Endowment. \\
\raggedright Proceedings of the VLDB Endowment. 
ISSN 2150-8097. \\
}\addtocounter{footnote}{-1}\endgroup

\ifdefempty{\vldbavailabilityurl}{}{
\vspace{.3cm}
\begingroup\small\noindent\raggedright\textbf{VLDB Workshop Artifact Availability:}\\
The source code, data, and/or other artifacts have been made available at \url{\vldbavailabilityurl}.
\endgroup
}

\section{Introduction}
Long-document analysis is an important but expensive task. Clinical records, legal contracts, and scientific papers are routinely examined in depth, and analyzing them by hand is slow and labor-intensive.  The expense grows further when the same document must be analyzed many times: a patient record is consulted by clinicians over time, a contract is revisited across many matters, and a scientific paper is queried for many different research questions. 
A recent and promising line of work uses large language models (LLMs) to do this automatically, answering the free-form questions that previous keyword and extraction tools could not~\cite{shankar2025docetl, patel2025lotus}.

When the LLM is prompted with the full document on each question, the cost of answering grows linearly with the size of the document. Long contexts add a problem of their own: as the input grows, the model fabricates answers more often at a rate that climbs steeply with context length, even when the answer is present in the document~\cite{liu2024lost}.
DocETL~\cite{shankar2025docetl}, together with related declarative pipelines such as Lotus~\cite{patel2025lotus} and Palimpzest~\cite{liu2025palimpzest}, makes the analysis more capable through agentic query rewriting and chunked execution. These systems are stateless: they maintain no state between queries, processing each question from scratch over the original document that result in two consequences follow. First, optimizing a single pass offers only limited acceleration. Second, the full cost recurs on every question, since nothing is amortized across them.

A natural optimization is to precompute some intermediate state once and reuse it across questions. Two families of approaches do this. \textbf{The first builds a vector index over the document.} Retrieval-augmented generation~\cite{lewis2020rag} splits the document into chunks and embeds each as a vector; GraphRAG~\cite{edge2024graphrag} and LightRAG~\cite{guo2025lightrag} add graph structure and community summaries on top for questions that span the whole document. The shared limitation is the vector index itself: splitting fragments connections between distant passages and top-$k$ retrieval surfaces only a few, so whole-document questions return incomplete context; the fixed-dimensional embedding is lossy; and the index is tied to one model, so it must be rebuilt whenever that model changes.
\textbf{The second family extracts chosen attributes into a structured table.} Information-extraction systems such as Evaporate~\cite{arora2023evaporate}, Doctopus~\cite{chai2025doctopus}, and ZenDB~\cite{lin2024zendb} do this; the result is precise for filter-style lookups, but cannot answer open-ended natural-language questions, and the attributes must be fixed before anyone knows what will be asked.

In this paper, we explore a different path. \textit{We return to the successful principles from the database literature, where answering many queries over a large body of data without scanning it is precisely the problem that the index, and in particular the B+ tree~\cite{comer1979ubiquitous, bayer1972organization}, has solved for decades.} We read each document once and compile it into a \textbf{hierarchical, plain-text index that any LLM can read directly}: a few high-level summaries at the root, topic-level summaries beneath them, and detailed page summaries at the leaves. The index is built once per document and consulted by every later analysis, exactly as a database index is built once and reused by every query.

This keeps the reuse benefit shared by vector indices and structured extraction while avoiding the weakness of each. Like a vector index, it is built once and reused, but because it is plain text it is model-independent and preserves the cross-document connections that chunking destroys. Like structured extraction, it imposes organization on the document, but because that organization is discovered by the LLM rather than fixed in a schema, it answers deep, open-ended questions rather than only filter-style lookups. Each node is a short summary sized to fit comfortably within the model's context window.

Although the index borrows the B+ tree's shape, hierarchical access over pages, it departs from it in three ways suited to LLM access. Content lives at every level, not only at the leaves, so many questions are answered from an internal summary without descending further, much as a covering index answers a query from the index alone without touching the base table~\cite{graefe2011btree}. The index is plain text rather than embedded vectors, so any model can read it and context is preserved by construction. Navigation is semantic: the model judges which summaries are relevant rather than comparing a search key, so the same index serves open-ended analysis rather than only exact-match or range queries.

In this work we explore the design of this new kind of index: how its structure is discovered from a corpus rather than hand-designed, how each document is compiled into it, and how questions are answered against it and the index refined over time. 

Section~\ref{sec:background} examines the limits of current practice in more detail. Section~\ref{sec:method} presents the index and its three design components. Section~\ref{sec:eval} reports a preliminary evaluation on NarrativeQA, and Section~\ref{sec:conclusion} concludes.

\section{Related Works}
\label{sec:background}

We group existing work on LLM-based long-document analysis into three families and point out where each is limited.

The first family answers each question by reading the document directly, keeping nothing from one question to the next. The simplest version feeds the whole text to the model along with the question. More developed pipelines decompose the work: DocETL~\cite{shankar2025docetl} breaks the analysis into a sequence of operators over chunks and uses agentic rewriting to improve accuracy, with follow-ups extending this line to interactive steering~\cite{shankar2025docwrangler} and multi-objective rewrites~\cite{wei2026multiobjective}; related declarative systems such as Lotus~\cite{patel2025lotus} and Palimpzest~\cite{liu2025palimpzest} optimize similar pipelines. These are flexible and need no setup, but reuse nothing across questions, so cost grows with each question and the long context buries relevant details, raising hallucination on long inputs~\cite{roig2026hallucinate}.

A second family avoids reprocessing by building reusable intermediate state. Retrieval-augmented generation~\cite{lewis2020rag} splits the document into chunks, embeds each as a vector, and at query time retrieves the chunks nearest the question. GraphRAG~\cite{edge2024graphrag} and LightRAG~\cite{guo2025lightrag} add structure over the chunks---an entity graph with community summaries in the former, a lighter dual-level graph in the latter---for broad questions that draw on many parts of the document at once. Closest to our design, RAPTOR~\cite{sarthi2024raptor} recursively clusters and summarizes chunks into a tree of summaries much like ours. All share two limitations: the index is tied to one embedding model, so changing either forces a rebuild; and chunking and top-$k$ retrieval are lossy, breaking links between distant passages and returning only a few chunks, so whole-document questions are easily missed. RAPTOR is no exception, since its tree is built bottom-up by clustering and queried by similarity, making its nodes retrieval targets rather than a plain-text schema that can be read and navigated directly.

A third family extracts chosen attributes into a structured table ahead of time, as in Evaporate~\cite{arora2023evaporate}, Doctopus~\cite{chai2025doctopus}, and ZenDB~\cite{lin2024zendb}. Querying is then a structured lookup, which is fast and precise for questions that can be written as filters or aggregations over known fields. However, there are two main limitations: the schema is usually fixed before the workload is known, so attributes that no one anticipated are simply absent; and a structured table can be a poor fit for free-form questions, such as ``why a character betrays a partner'', which call for synthesis across the document rather than simple value lookups.
\section{Method}
\label{sec:method}

\subsection{Overview}
\label{sec:overview}

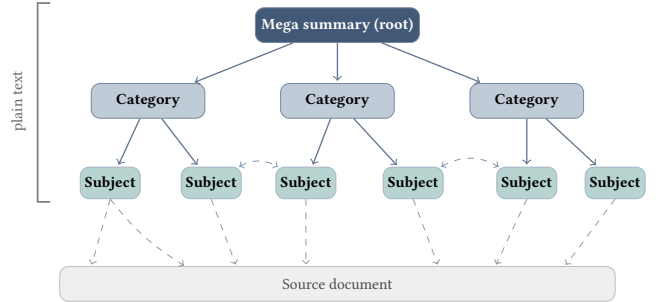
\begin{figure}[t]
  \centering
  \definecolor{navy}{HTML}{1F3A5F}
  \definecolor{midnavy}{HTML}{4A6B8F}
  \definecolor{softteal}{HTML}{B8D4D1}
  \definecolor{paleteal}{HTML}{DCE9E7}
  \definecolor{softgray}{HTML}{EFEFEF}
  \resizebox{\columnwidth}{!}{%
  \begin{tikzpicture}[
      every node/.style={font=\footnotesize},
      box/.style={draw=navy!60, line width=0.4pt, rounded corners, align=center,
                  font=\footnotesize\bfseries, minimum height=0.55cm, text=black,
                  inner sep=2pt},
      leaf/.style={draw=midnavy!50, line width=0.4pt, rounded corners,
                   fill=softteal, align=center, font=\footnotesize\bfseries,
                   minimum height=0.5cm, minimum width=0.95cm, text=black,
                   inner sep=2pt},
      src/.style={draw=black!25, line width=0.4pt, rounded corners,
                  align=center, font=\footnotesize, fill=softgray,
                  minimum height=0.55cm, text=black!70, inner sep=2pt},
      hier/.style={->, semithick, draw=navy!70, shorten >=1pt},
      prov/.style={->, dashed, draw=black!40, shorten >=1pt},
      cross/.style={<->, dashed, draw=midnavy!70}
    ]
    \node[box, fill=navy!85, text=white, minimum width=2.6cm] (mega) at (0,2.6) {Mega summary (root)};
    \node[box, fill=midnavy!35, minimum width=1.8cm] (s1) at (-3.0,1.4) {Category};
    \node[box, fill=midnavy!35, minimum width=1.8cm] (s2) at (0,1.4) {Category};
    \node[box, fill=midnavy!35, minimum width=1.8cm] (s3) at (3.0,1.4) {Category};
    \node[leaf] (p1) at (-3.6,0.1) {Subject};
    \node[leaf] (p2) at (-2.0,0.1) {Subject};
    \node[leaf] (p3) at (-0.5,0.1) {Subject};
    \node[leaf] (p4) at (1.2,0.1) {Subject};
    \node[leaf] (p5) at (3.0,0.1) {Subject};
    \node[leaf] (p6) at (4.4,0.1) {Subject};
    \node[src, minimum width=8.8cm] (src) at (0,-1.5) {Source document};

    \draw[hier] (mega) -- (s1);
    \draw[hier] (mega) -- (s2);
    \draw[hier] (mega) -- (s3);
    \draw[hier] (s1) -- (p1);
    \draw[hier] (s1) -- (p2);
    \draw[hier] (s2) -- (p3);
    \draw[hier] (s2) -- (p4);
    \draw[hier] (s3) -- (p5);
    \draw[hier] (s3) -- (p6);

    \draw[prov] (p1.south) -- (-3.9,-1.225);
    \draw[prov] (p1.south) to[bend right=10] (-2.4,-1.225);
    \draw[prov] (p2.south) -- (-1.6,-1.225);
    \draw[prov] (p3.south) -- (-0.5,-1.225);
    \draw[prov] (p4.south) -- (1.6,-1.225);
    \draw[prov] (p5.south) -- (2.5,-1.225);
    \draw[prov] (p6.south) -- (3.6,-1.225);

    \draw[cross] (p2) to[bend left=30] (p3);
    \draw[cross] (p4) to[bend left=30] (p5);

    \draw[thick, draw=black!50] (-4.55,2.95) -- (-4.75,2.95) -- (-4.75,-0.20) -- (-4.55,-0.20);
    \node[rotate=90, font=\footnotesize, text=black!60] at (-5.05,1.4) {plain text};
  \end{tikzpicture}%
  }%
  \caption{Proposed hierarchical plain-text index. A query enters at the root and follows the relevant pages, moving down toward specifics or hopping across to related subjects. Each subject links back to one or more spans in the source document.}
  \label{fig:index}
\end{figure}

Our proposed index is a hierarchical plain-text structure compiled once per document (Figure~\ref{fig:index}). At the top, a \emph{root} (the mega summary) gives an overview of the document and points to the broad areas beneath it, called \emph{categories}. A category may stand on its own or split into sub-categories, so the index has one or more layers depending on how the document divides. Under the lowest layer sit \emph{subjects}, the specifics, such as a character or an event in a screenplay, each recording a small set of \emph{fields} (the details about that particular character or event) filled in directly from the document. Every page, root, category, or subject, is plain text and small enough to fit within an LLM's context window.

The remaining sections work through the three design problems: how the structure is decided (Section~\ref{sec:discovery}), how a document is compiled into it (Section~\ref{sec:compile}), and how questions are answered against it (Section~\ref{sec:query}).

\subsection{Discovering the structure}
\label{sec:discovery}

For each source document, the LLM reads it and proposes a \emph{schema} for the index: a hierarchical structure of categories and fields. At the top, a brief overview known as the \emph{mega summary} sits at the root. Beneath it, the LLM organizes the document into categories, and under each category, into subjects with a small set of fields (the kinds of details to record about each entry). On a movie screenplay, for example, the schema might propose a movie-wide overview at the root, with categories \textit{characters}, \textit{events}, and \textit{facts} beneath it; subjects under \textit{characters} would record fields such as \textit{role}, \textit{actions}, and \textit{relationships}. We do not dictate which categories or fields the model should produce; both are proposed by the model from what it sees in the document.

The subjects and the values that fill their fields are left to be determined during compilation (Section~\ref{sec:compile}). In the current implementation, we keep each subject small with a fixed cap per field. A more principled mechanism would bound the number of links at each layer, analogous to the fanout of a B+ tree, so lookup stays efficient as the index grows; designing such a mechanism is a direction we leave to future work.

\subsection{Compiling a document}
\label{sec:compile}

Compilation reads the document in full and fills in the schema from discovery. It runs in three steps: clean, extract, and write.

\medskip
\noindent\textbf{Cleaning.} The raw text is stripped of download artifacts and markup, then split into chunks of about 30{,}000 characters that overlap by 2{,}000, so that a passage at the edge of one chunk also appears in full in the next.

\medskip
\noindent\textbf{Extraction.} The schema from discovery is handed to the LLM as part of the extraction prompt: it tells the model which categories to look for and which fields to fill in under each. For every chunk, the model reads the text against this schema and returns the subjects it finds (the characters, events, and facts that fit each category) with their fields filled in. The categories themselves never change; only the list of subjects under each one grows. For example, in a screenplay, if chunk 1 mentions two events, the \textit{events} category gains two subjects; if chunk 2 mentions one more, it gains another. We call the running record of all categories and the subjects under them the \emph{merged record}. For populating fields, the first non-empty answer is kept; if a later chunk supplies a different one, both are retained so nothing the document says is quietly lost. Each extracted field is also tagged with the source offset of the passage that produced it, so any value on a subject can be traced back to the original text.

\medskip
\noindent\textbf{Writing.} Writing produces one file per subject grouped under its category, and a single overview file at the root summarizing the schema, matching the hierarchy in Figure~\ref{fig:index}.

\subsection{Answering queries}
\label{sec:query}

To answer a question, the system navigates the index. Each subject carries a short summary; navigation scores these summaries against the question to decide where to go next. It starts at the mega summary (root) and follows the most relevant link, descending into categories, jumping across to related subjects, or stepping back up when an area is exhausted.

Descending into the index is unbounded: the system may go as deep along a promising path as the index allows and visit neighboring pages. What is bounded is the number of times it may level up to revisit a higher node and then descend into a different branch, so traversal cannot wander indefinitely across the index. We currently set this branching budget to a fixed value, the same for every question; deciding it more intelligently, for instance from the complexity of the question or the size of the index, is left to future work. When the pages reached do not contain the answer, the system falls back to the most relevant part of the source document and answers from there. The index is not updated on a miss; folding fallback results back into the index is a direction we touch on in Section~\ref{sec:conclusion}.

\section{Preliminary Evaluation}
\label{sec:eval}
 
\subsection{Experimental Setup}

We evaluate on the NarrativeQA dataset~\cite{kocisky2018narrativeqa}, using ten movie screenplays and the 295 questions paired with them; the screenplays run to hundreds of pages and the questions are free-form, which is close to the setting we target. GPT-5 serves as both the compiler and the question-answering model, and an LLM judge scores answer equivalence against the reference answers.We compare against a full-document baseline (the entire script fed to the model on each question), the DocETL pipeline~\cite{shankar2025docetl}, and three retrieval baselines (RAG~\cite{lewis2020rag}, GraphRAG~\cite{edge2024graphrag}, and RAPTOR~\cite{sarthi2024raptor}).

We report two measures. \textbf{Accuracy} is the fraction of answers the LLM judges equivalent to the reference. \textbf{Cost} is the total number of input tokens a system consumes, which we separate into a one-time index build cost and the recurring cost of answering each question. All numbers are measured from a complete, automated run of each system, with no manual intervention or hand-tuned estimates.

\subsection{Results}
\begin{figure}[t]
  \centering
  \includegraphics[height=3.5cm,keepaspectratio]{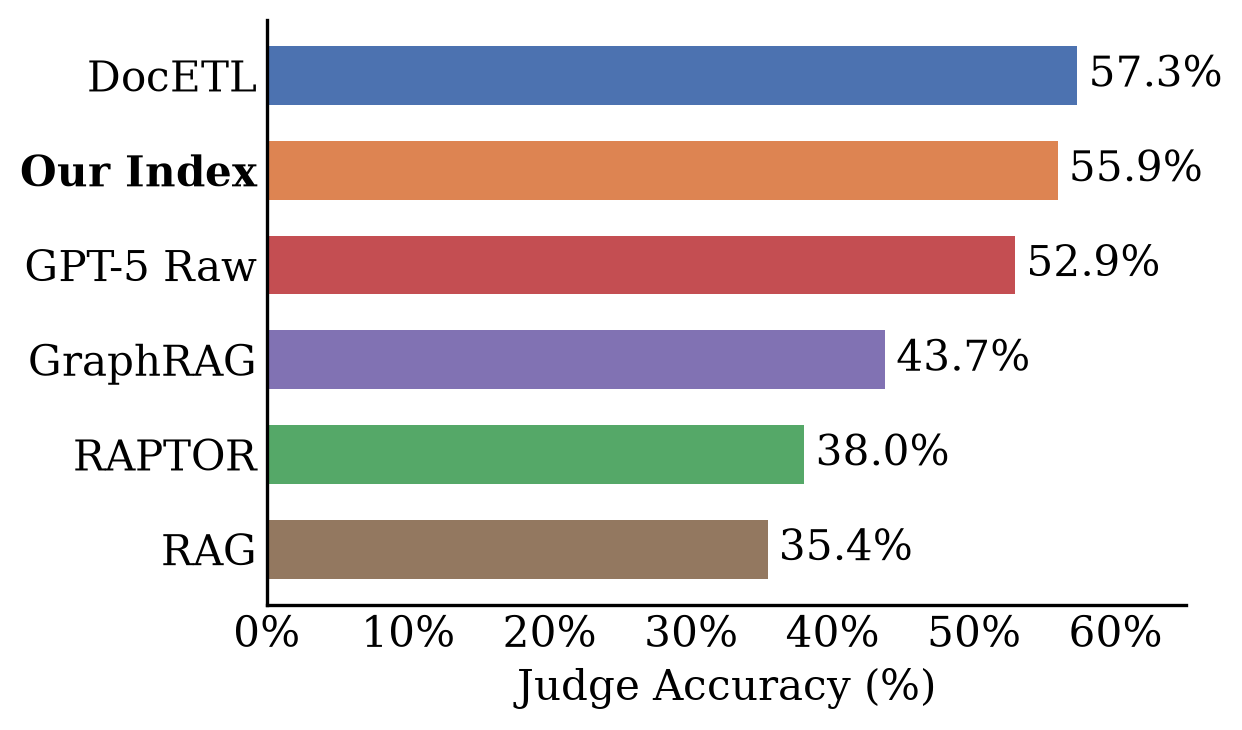}
  \caption{Judge accuracy on NarrativeQA. Our index is close to the DocETL pipeline and ahead of the retrieval baselines.}
  \label{fig:accuracy}
\end{figure}
\begin{figure}[t]
  \centering
  \includegraphics[height=3.5cm,keepaspectratio]{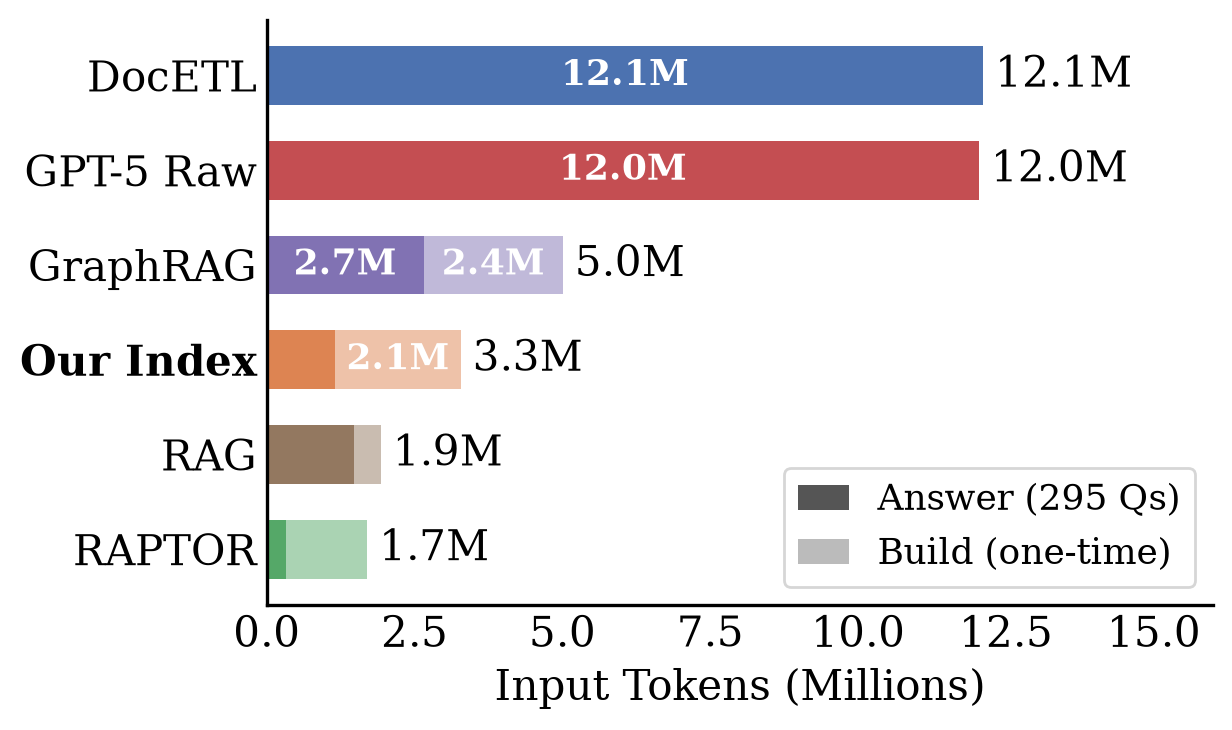}
  \caption{Total input-token cost of each system, split into one-time build (faded) and recurring per-query-batch answer (solid), for 295 questions over ten movies. Our index pays a one-time build cost and a low marginal answer cost.}
  \label{fig:cost}
\end{figure}
Figure~\ref{fig:accuracy} gives the accuracy ranking. Our index reaches 55.9\% judge accuracy, essentially matching DocETL, the strongest baseline, at 57.3\%; the gap is about one and a half points, roughly four questions out of 295. It also outperforms feeding the full script to the model (52.9\%). The three retrieval baselines trail: GraphRAG reaches 43.7\%, RAPTOR 38.0\%, and RAG 35.4\%. They struggle on long screenplays because they keep only a small slice of the document. RAG's chunked retrieval returns just a few passages per question, and GraphRAG's extractor produces a sparse graph (about 63 nodes per film with 88\% of them disconnected), so the parts of the document a free-form question depends on are usually missing. For RAPTOR we apply our system's strict rule: it must return \emph{not found} when the retrieved nodes or top-$k$ chunks lack the answer, rather than guessing from context, so all systems are judged on equal footing.

\begin{figure}[t]
  \centering
  \includegraphics[height=3.5cm,keepaspectratio]{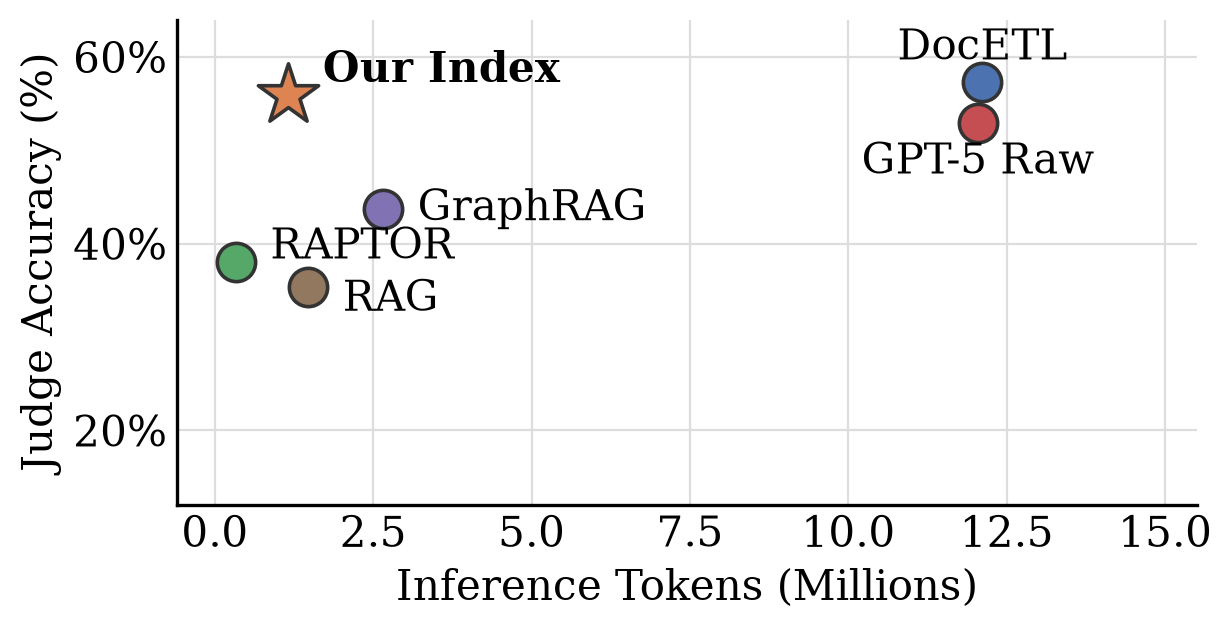}
  \caption{Answer cost versus judge accuracy on NarrativeQA (ten screenplays, 295 questions). Our index (star) reaches accuracy close to the most accurate baseline at a lower cost.}
  \label{fig:tradeoff}
\end{figure}

Figure~\ref{fig:cost} breaks each system's cost into the recurring per-query answer (solid) and the one-time build (faded). Our index spends 2.1M tokens to build the index once and 1.2M to answer all 295 questions, a marginal cost of roughly 3.9K tokens per question. DocETL and GPT baselines carry no build cost, but their entire 12M tokens is recurring, about 41K tokens per question for DocETL. Because the build cost is paid once and amortized across every later question, our total of 3.3M is about 73\% below DocETL's 12.1M and 73\% below the full-document baseline's 12.0M.
Figure~\ref{fig:tradeoff} places each system on the answer cost vs. accuracy plane; we count only inference tokens here and exclude the one-time build cost so per-query economics are directly comparable. Our index sits near the top of the accuracy range while spending fewer inference tokens than any other system: at 1.2M it answers all 295 questions for less than even the retrieval baselines, including RAPTOR's 1.7M, while reaching nearly the same accuracy as DocETL at a fraction of its per-question cost.
\section{Conclusion and Future Work}
\label{sec:conclusion}
 
We explored an indexing-based solution for long-document analysis and proposed a compiled, hierarchical, plain-text index: built once per document, organized like a B+ tree but with content at every level, with its structure discovered from the corpus rather than hand-designed. On NarrativeQA, the index matches the accuracy of the strongest stateless baselines at 40\% lower cost, while remaining model-independent and able to answer open-ended questions. These results are preliminary and confined to narrative text, but they suggest the design is worth pursuing.
 
Currently, we keep pages small with fixed caps on how much each page records, the same for every document, and we do not bound the number of pages or balance the hierarchy. A more principled scheme would derive these limits from a single constraint, how much context the question-answering model can use before its accuracy degrades on long inputs~\cite{liu2024lost}.
The index would then be built by a single recursive rule and fit a node's content on one page if it can. Otherwise the node should split into children and apply the same rule to each, so the depth grows only as far as the content forces rather than being fixed in advance. This raises further questions we want to study: how many layers a document of a given size should induce, how to choose the summary at each internal node so it guides navigation without duplicating the pages below it, and how to bound the children of a node so lookup stays efficient as the index grows. A separate direction is whether the index should change while it is being queried: today it is fixed once compiled and simply reports a miss when a question falls outside it, but a more capable system would re-read the source on a miss, fold what it finds back into the index, and eventually grow new structure toward the questions actually being asked, without bloating back into a copy of the full document. The work we are undertaking next is to solve these questions, while validating the approach across more domains and recalibrating the judge for each.
 

\bibliographystyle{ACM-Reference-Format}
\bibliography{sample}

\end{document}